\documentclass[10pt,twocolumn,english,prd,superscriptaddress,nofootinbib,preprintnumbers,showpacs,floatfix]{revtex4}

\usepackage[utf8]{inputenc}
\usepackage{doi}
\usepackage{hyperref}
\hypersetup{
  colorlinks=true,        
  linkcolor=blue,         
  citecolor=cyan,         
}
\usepackage{graphicx}
\usepackage{bm}
\usepackage{latexsym}
\usepackage{dcolumn}
\usepackage{amsmath,amsfonts,amssymb}
\usepackage{physics}
\usepackage{amsthm}
\usepackage{color}
\usepackage{xcolor}
\usepackage{graphicx,float}
\usepackage{comment}
\usepackage{hyperref}
\usepackage[normalem]{ulem}
\usepackage{enumitem} 
\usepackage{ulem}

\usepackage{orcidlink}

\usepackage{pdfpages}
\usepackage{epstopdf}
\usepackage[utf8]{inputenc}

\def\be {\begin{equation}}
\def\ee {\end{equation}}
\def\bea {\begin{eqnarray}}
\def\eea {\end{eqnarray}}
\def\bc {\begin{center}}
\def\ec {\end{center}}
\def\bfg {\begin{figure}}
\def\efg {\end{figure}}
\def\bi {\begin{itemize}}
\def\ei {\end{itemize}}

\def\beq{\begin{equation}}
\def\eeq{\end{equation}}
\def\br{\begin{eqnarray}}
\def\er{\end{eqnarray}}
\newcommand{\eel}[1] {\label{#1}\end{equation}}

\newcommand{\bdm}{\begin{displaymath}}
\newcommand{\edm}{\end{displaymath}}

\usepackage{orcidlink}

\begin{document}
\title{Emergence of ${\rm ER}={\rm EPR}$ from non-local gravitational energy}

\author{Kimet Jusufi\,\orcidlink{0000-0003-0527-4177}}
\email{kimet.jusufi@unite.edu.mk}
\affiliation{Physics Department, State University of Tetovo, Ilinden Street nn, 1200, Tetovo, North Macedonia}

\author{Francisco S.N. Lobo\,\orcidlink{0000-0001-9155-7282}}%
\email{fslobo@fc.ul.pt}
\affiliation{Departamento de F\'{i}sica, Faculdade de Ci\^{e}ncias da Universidade de Lisboa, Campo Grande, Edif\'{\i}cio C8, P-1749-016 Lisbon, Portugal}
\affiliation{Instituto de Astrof\'{\i}sica e Ci\^{e}ncias do Espaco, Faculdade de Ci\^encias da Universidade de Lisboa, Campo Grande, Edif\'{\i}cio C8, P-1749-016 Lisbon, Portugal}%

\author{Emmanuel N. Saridakis\,\orcidlink{0000-0003-1500-0874}} 
\email{msaridak@noa.gr}
\affiliation{National Observatory of Athens, Lofos Nymfon, 11852 Athens, Greece}
\affiliation{Departamento de Matem\'{a}ticas, Universidad Cat\'{o}lica del 
Norte, Avda. Angamos 0610, Casilla 1280 Antofagasta, Chile}
\affiliation{CAS Key Laboratory for Researches in Galaxies and Cosmology, 
School 
of Astronomy and Space Science, University of Science and Technology of China, 
Hefei, Anhui 230026, China}

\author{Douglas Singleton\,\orcidlink{0000-0001-9155-7282}}
\email{dougs@mail.fresnostate.edu}
\affiliation{Physics Department, California State University, Fresno, CA 93740}

\begin{abstract}

We construct a class of wormhole geometries supported by the non-local gravitational self-energy that regularizes the particle and black-hole sectors of spacetime. Using this framework, inspired by T-duality, we show that two entangled particles (or particle–black-hole pairs) naturally source an Einstein-Rosen–type geometry in which the required violation of the strong energy condition arises from intrinsic quantum-gravity effects rather than from ad hoc exotic matter, which is matter that violates the null energy condition. We classify the resulting wormholes, analyze their horizons, throat structure and embedding properties, and we identify the exotic energy needed at the minimal surface. Imposing the ${\rm ER}={\rm EPR}$ requirement of non-traversability and the absence of a macroscopic throat, we find that only the zero-throat geometry is compatible with an entanglement-induced Einstein–Rosen bridge, providing a concrete realization of ${\rm ER}={\rm EPR}$ within a fully regular spacetime. Finally, we briefly discuss possible implications for microscopic ER networks from vacuum fluctuations, replica-wormhole interpretations of Hawking radiation, and possible links to entanglement-driven dark-energy scenarios.

\end{abstract}

\pacs{  04.20.Gz, 04.60.Bc,  03.65.Ud}
\maketitle

\section{Introduction}

A geometric precursor to wormholes was noted by Flamm in 1916~\cite{2015GReGr..47...72F}, who observed that the spatial slice of the Schwarzschild solution can be visualized as a curved surface linking two asymptotically flat regions. Nearly two decades later, Einstein and Rosen made this connection explicit by identifying a minimal surface joining two identical sheets, introducing what is now known as the Einstein–Rosen bridge~\cite{einstein1935particle}. Since then, a wide range of classical and quantum–gravity extensions of wormhole physics have been explored \cite{Wheeler:1955zz,Misner:1960zz,Ellis:1973yv,Bronnikov:1973fh,Morris:1988cz,Morris:1988tu,Visser:1989kh,Visser:1989kg,Echeverria:1991nk,Visser:1992tx,Ford:1995wg,Hochberg:1997wp,Kim:2001ri,Shinkai:2002gv,Hochberg:1996ee,Lobo:2004rp,Boehmer:2007md,Lobo:2008zu,Harko:2008vy,Harko:2009xf,Bronnikov:2010tt,Garcia:2010xb,Boehmer:2012uyw, Lobo:2012qq,Harko:2013yb,Simpson:2018tsi,Dai:2019mse,Bronnikov:2021liv,DeFalco:2021ksd},  driven in large part by advances in quantum gravity, quantum information theory, and holography \cite{Hawking:1987mz,Garfinkle:1990eq,Jensen:2013ora,Lobo:2013vga,Lobo:2014nwa,maldacena2016remarks,maldacena2021syk,Capozziello:2018mqy,Maldacena:2018lmt,kitaev2015simple,dai2020testing,Jafferis:2022crx,Jafferis2022,ali2022unitary,Jusufi:2023dix,Tsilioukas:2023tdw,Licata:2025rik,Hadi:2025zoe,Liu:2025zts,Lobo:2025nng,Lobo:2014fma}. This has motivated the search for geometries capable of realizing ER bridges without singularities or exotic matter in the traditional sense.

A major conceptual development came with the ${\rm ER}={\rm EPR}$ conjecture of Maldacena and Susskind \cite{maldacena2013cool}, which proposes a correspondence between Einstein--Rosen (ER) bridges \cite{einstein1935particle} and the Einstein--Podolsky--Rosen (EPR) correlations underlying quantum entanglement \cite{bengtsson2017geometry,einstein1935can}. This idea suggests that spacetime connectivity and quantum entanglement may reflect two aspects of a single underlying structure.
Despite its appeal, no generally accepted geometry has been identified that realizes an ER bridge compatible with the particle sector. In this work we construct a regular ER-like geometry from non-local gravitational self-energy and classify which wormholes satisfy ER=EPR.

The original Einstein--Rosen construction is derived from the Schwarzschild metric, which contains a curvature singularity and therefore cannot represent a nonsingular particle-like configuration. It is instead applicable only to macroscopic gravitational systems such as black holes, particularly supermassive ones. In addition, as shown in \cite{Guendelman:2009er,Guendelman:2009zz}, the Einstein--Rosen bridge requires exotic matter that violates the null energy conditions \cite{Morris:1988cz,Visser:1995cc,Lobo:2017cay}, and such matter cannot arise within standard General Relativity (GR). This limitation implies that the classical Einstein--Rosen wormhole cannot be realized as a stable or traversable configuration. Thus, this motivates the study of alternative geometries or extensions of GR capable of supporting wormhole solutions without the need for exotic energy sources.

In the literature one can study a specific gravitational potential arising from string T-duality \cite{Sathiapalan:1986zb}, in which both the potential and the resulting spacetime geometry are manifestly free of singularities. String T-duality relates string theories compactified on higher-dimensional spaces of radius $R$ to theories compactified on radii ${R^\star}^2 / R$, where $R^\star = \sqrt{\alpha'}$ denotes the self-dual radius associated with the Regge slope $\alpha'$. Under this duality transformation, the Kaluza--Klein excitation number $n$ and the winding number $w$ are interchanged, reflecting the symmetry between momentum and winding modes in closed string spectra. The corresponding quantum-corrected propagator remains regular at all scales and naturally incorporates an ultraviolet cutoff set by the zero-point length $l_{0} = 2L$ \cite{Padmanabhan:1996ap,Smailagic:2003hm,Fontanini:2005ik}. This short-distance regulator is a characteristic feature of T-duality–inspired effective models and ensures the absence of curvature singularities.
Such T-duality–motivated frameworks have been extensively explored in black hole and cosmological contexts \cite{Nicolini:2022rlz,Nicolini:2019irw,Gaete:2022ukm,Nicolini:2023hub,Jusufi:2024dtr,Jusufi:2025qgd,Jusufi:2022mir,Millano:2023ahb}, where they provide a robust mechanism capable of resolving singularities in both neutral and charged black hole solutions, yielding fully regular geometries consistent with the underlying string-theoretic structure.
Hence, in the following analysis  T-duality is employed purely as a guiding heuristic for introducing a minimal length and a regular core, rather than as a direct result derived from string theory.

In this work we are interested in  constructing a wormhole connecting two entangled states by employing a regular spacetime geometry capable of describing both the particle sector and the black hole sector, including Planck-mass particle–black hole configurations. In this framework, the gravitational self-energy plays a central role in determining the geometry and its physical interpretation. The appearance of exotic matter at the wormhole throat, required to satisfy the junction conditions, is directly associated with the violation of classical energy conditions. In regular spacetimes, such violations arise naturally from quantum gravity effects rather than from ad hoc stress–energy sources. Consequently, this setting provides a consistent mechanism through which quantum-induced energy-condition violations may trigger a smooth transition between black hole and wormhole phases for two entangled states.

A central result of this work is that, among all wormhole geometries supported by the non-local gravitational energy, only the zero-throat configuration satisfies the causal and geometric requirements of the ${\rm ER}={\rm EPR}$ correspondence. As we will show, the non-local energy uniquely selects the zero-throat, non-traversable wormhole as the only geometry compatible with  ${\rm ER}={\rm EPR}$, thereby providing a concrete entanglement-induced realization of an Einstein–Rosen bridge in a fully regular spacetime.

The paper is organized as follows. In Section \ref{SecII}  we review the non-local gravitational self-energy and the resulting regular spacetime geometry that unifies the particle and black-hole sectors. In Section \ref{SecIII} we employ this framework to construct the full family of Einstein–Rosen–type wormhole geometries sourced by two entangled subsystems, analyzing their horizons, throat structure, embedding properties, and the exotic matter required at the minimal surface. In Section \ref{SecIV}  we examine the physical consistency of these geometries within the ${\rm ER}={\rm EPR}$ correspondence, identifying which solutions admit an interpretation as entanglement-induced Einstein–Rosen bridges and discussing broader implications for quantum gravity. Finally, Section \ref{SecV} contains our conclusions.

\section{Regular spacetime from non-local gravitational energy}
\label{SecII}

In this section we briefly review the basics of non-local gravitational energy and the corresponding 
regular spacetime geometry, following  \cite{Jusufi:2025qgd}.

\subsection{Non-local gravitational self energy}  

We start by recalling the quantum-corrected static interaction potential derived from the field theory with the path integral duality proposed in \cite{Nicolini:2019irw}. 
Specifically, the momentum-space massive propagator induced by the path integral duality is written as \cite{Nicolini:2022rlz,Nicolini:2023hub}
\begin{equation}
G(k)
= -\frac{l_0}{\sqrt{k^2+m_0^2}}\, K_1 \!\left(l_0 \sqrt{k^2+m_0^2}\right),
\end{equation}
with $l_0$ being the zero-point length and $K_1\!\left(x\right)$ is a modified Bessel function of the second kind. Without loss of generality, we consider the massless propagator case, i.e. $m_0=0$. 
Specifically, we have two cases: at small momenta, we obtain the conventional massless propagator $G(k) = -k^{-2}$. At large momenta, the exponential suppression is responsible for curing UV divergences \cite{Nicolini:2022rlz,Nicolini:2023hub}. Consider a static external source $J$ which consists of two point-like masses, $m$ and $M$, at relative distance $\vec{r}$, then by means of the generating
functional of connected diagrams $W[J]$ that equals the integral of the interaction energy over the time span $T$ one can find the potential as \cite{Nicolini:2019irw}
\begin{eqnarray}\notag
V_G(r) &=& -\frac{1}{m}\frac{W[J]}{T} \\\notag
&=& -M\, \int \frac{d^3 k}{{\left(2\pi\right)}^3}\; { G_{F}(k)|}_{k^0=0}\; 
 \exp\!\left(i \vec{k} \vec{r}\right) \\
 &=&
 -\frac{M}{\sqrt{r^2 + l_0^2}}. \label{potential}
\end{eqnarray}
Using Poisson's equation gives an energy density function for the bare matter as
\cite{Nicolini:2019irw}
\begin{equation}
\rho^{\rm bare}(r)= \frac{1}{4\pi }\nabla^2 V_G(r)=\frac{3 l_0^2 M}{4 \pi \left( r^2+l_0^2\right)^{5/2}}.\label{density}
\end{equation}

In contrast to general relativity, which is a local theory, here we aim to incorporate an additional term (i.e., the self-energy associated with the gravitational field) that arises as a non-local effect, leading to a new contribution in the field equations. 
Specifically, we consider a modified, non-local theory of gravity recently proposed in Ref.~\cite{Jusufi:2025qgd}, where it was shown that a non-local gravitational effect can be deduced by assuming
 \begin{align}
\rho^{\text{GSE}}(\textbf{x}) =\int  \mathcal{R} (\textbf{x} -\textbf{y} ) \rho^{\text{bare}}(\textbf{y} ) d^3 \textbf{y},
\end{align}
 where GSE stands for ``gravitational self energy'', and with 
 $\mathcal{R} $  the spatial kernel. Note that  this expression is local in time and non-local in space. In addition, we have \cite{Jusufi:2025qgd}
\begin{equation}
\mathcal{R} (\textbf{x}-\textbf{y}) = \frac{\mathcal{C} (\textbf{x})}{\sqrt{|\textbf{x}-\textbf{y}|^2+l_0^2}} ~,
\end{equation}
where $\mathcal{C} (\textbf{x})$ is some parameter with appropriate units. We can set $\mathcal{C} (\textbf{x})=\rho^{\rm bare} (\textbf{x})$. By neglecting the dimensions, one can write $\rho^{\rm bare}=M \delta(\textbf{x})$. Under this choice and since we are interested in the static case, there are no dynamical new equations for the matter. 
We can compute the average self-energy using the fact that 
 \begin{equation}
 E^{\text{GSE}}  :=\frac{1}{2}\int \rho^{\text{GSE}}(\textbf{x})\, d^3 \textbf{x},
 \end{equation}
and by using the relation for the gravitational potential
 \begin{equation}
V_G(\textbf{x}) = - \int \, \frac{\rho^{\text{bare}}(\textbf{y})}{\sqrt{|\textbf{x}-\textbf{y}|^2+l_0^2}}\, d^3 \textbf{y}~,
 \end{equation}
it follows that \cite{Jusufi:2025qgd}
 \begin{equation}
 E^{\text{GSE}} = -\frac{1}{2} \int V_G(\textbf{x})\, \rho^{\text{bare}}(\textbf{x})\, d^3 \textbf{x}.
 \end{equation}
Using Gauss's law for gravity $\nabla \cdot \mathbf{g} = -4\pi \rho^{\text{bare}}$, and going to the Newtonian limit, we can compute the nonlocal gravitational self-energy as
\begin{equation}
 E^{\text{GSE}}  =
- \frac{1}{8\pi} \int \mathbf{g} \cdot (\nabla V_G)\, d^3\mathbf{r}. 
\end{equation}
For the distant observer, only this term gives a contribution. To obtain the gravitational self-energy term to the energy-momentum tensor we will use a coordinate-independent quantity observable by an asymptotic observer in the Newtonian limit, namely the Newtonian gravitation field defined as 
\begin{eqnarray}
    \textbf{g}=-\nabla V_G(r)= -\frac{Mr}{(r^2 + l_0^2)^{3/2}}\hat{r}.
\end{eqnarray}
The gravitational self-energy term is then proportional to the volume integral of the square of $\textbf{g}$ \cite{Jusufi:2025qgd}
\begin{eqnarray}\label{GSE}
 E^{\rm GSE}  &=& \frac{1}{8\pi}\int_0^r \textbf{g}^2(r') d^3\textbf{r}' \nonumber \\
&=&-\frac{5 M^2 r^3}{16(r^2+l_0^2)^2}-\frac{3 M^2 l_0^2 r}{16(r^2+l_0^2)^2}\\
&&+\frac{3  M^2}{16 l_0} \arctan(\frac{r}{l_0}). \nonumber
\end{eqnarray}

This result plays an important role in the spacetime geometry.

\subsection{Regular spacetime geometry}

Let us now review the solution of the spacetime geometry presented in \cite{Jusufi:2025qgd}.
We impose a spherically symmetric metric of the form
\begin{equation}
    ds^2=-f(r)dt^2+f(r)^{-1}dr^2+r^2(d\theta^2+\sin^2\theta d\phi^2),
\end{equation}
with 
\begin{eqnarray}
    f(r)=1-\frac{2 m(r)}{r}.
\end{eqnarray}
By using the conservation of the energy-momentum tensor $ \nabla_\mu  \mathcal{T}^{\mu \nu}=0,$
we can obtain \cite{Jusufi:2025qgd}       the condition $\rho=-\mathcal{P}_r$, and then 
it follows that
\begin{eqnarray}
   \mathcal{P}_T=-\rho-\frac{r}{2} \frac{d \rho}{dr}.
\end{eqnarray}

In general, for the total energy-momentum tensor  one can write  
the following relation \cite{Jusufi:2025qgd}
\begin{eqnarray}
    \mathcal{T}_{\mu \nu}=-\rho  g_{\mu \nu}+\tau_{\mu \nu} \ ,
\end{eqnarray}
where $\rho$ is now the total energy density given by \cite{Jusufi:2025qgd}
\begin{eqnarray}
    \rho=\rho^{\rm bare}+\rho^{\rm GSE},\,\,\, \text{and}\,\,\,\,\tau_{\mu \nu}&=& \tau_{\mu \nu}^{\rm bare}+ \tau_{\mu \nu}^{\rm GSE},
\end{eqnarray}
where
\begin{eqnarray}
    \tau_{\mu \nu}&=&{\rm diag} \left(0, 0, -(r/2)\, (\partial_r \rho), -(r/2)\, (\partial_r \rho)  \right).
\end{eqnarray} 

Now, the Poisson's equation gives an energy density function for 
the bare matter and the energy density. In Eq. \eqref{density} we obtained the radial ($P_r$) and 
transverse ($P_T$) pressure   as \cite{Jusufi:2025qgd}
\begin{eqnarray}
    P_r^{\rm bare}&=&-\rho^{\rm bare}
    =-\frac{3 l_0^2 M}{4 \pi \left( r^2+l_0^2\right)^{5/2}},\\
    P_T^{\rm bare}&=&\frac{3 l_0^2 M (3r^2-2 l_0^2)}{8 \pi \left( r^2+l_0^2\right)^{7/2}}.
\end{eqnarray}

An important point to be mentioned is that by using the bare matter energy-momentum tensor it can be seen that the Strong
Energy Condition (SEC) ({\it i.e.} $\rho^{\rm bare}+\sum_i P^{\rm bare}_i \geq 0$) is violated in the region $r < \sqrt{2/3} \,l_0$ \cite{Jusufi:2025qgd}. 
One can also obtain the pressure components of the GSE energy-momentum tensor as
\begin{equation}\label{Pressurer}
   P^{\rm GSE}_r=-\rho^{\rm GSE}
    =-\frac{M^2r^2}{8\pi(r^2 + l_0^2)^3}
\end{equation}
and
\begin{eqnarray}
    P_T^{\rm GSE}&=&\frac{r^2 M^2 (r^2-2 l_0^2)}{8 \pi \left( r^2+l_0^2\right)^4}.
\end{eqnarray}

Similarly, for the GSE energy-momentum tensor, one can see that the SEC ({\it i.e.} $\rho^{\rm GSE}+\sum_i P^{\rm GSE}_i \geq 0$) is again violated in the region $r < \sqrt{2} \,l_0$ \cite{Jusufi:2025qgd}. In the following, we will see that such a violation of energy conditions can play an important role in the wormhole stability, which requires exotic matter at the throat.  

Furthermore, we can assume that the total mass of the system is given by the bare mass plus the mass/energy stored in the gravitational field. Specifically, we can compute the mass enclosed in some region by using the energy density for the bare matter and non-local gravitational energy, namely  \cite{Jusufi:2025qgd}
\begin{eqnarray}
    m(r)=4 \pi \int_0^r \left[\rho^{\rm bare}(r') +\rho^{\rm GSE}(r') \right]r'^2 dr'.
\end{eqnarray}
{\it  This mass represents the bare mass plus the contribution from smeared gravitational self-energy.}
In this way, by solving the above integral for the metric function, we   extract the result \cite{Jusufi:2025qgd}
\begin{eqnarray}
    f(r)=1-\frac{2  M r^2}{ (r^2+l_0^2)^{3/2}}+\frac{  M^2 r^2  }{(r^2+l_0^2)^2} F(r)\label{metric1},
\end{eqnarray}
where 
\begin{equation}\label{F}
    F(r)=\frac{5}{8}+\frac{3l_0^2}{8 r^2}-\frac{3 (r^2+l_0^2)^2}{8 l_0 r^3}\arctan(\frac{r}{l_0}).
\end{equation} 
As shown in  \cite{Jusufi:2025qgd}, depending on the value of the mass parameter, we can have:\\ 

(i) The \textit{particle sector}, if the mass belongs to the region   $M\ll M_{Pl}$. \\

(ii) The \textit{particle-black hole sector}, if the mass is of Planck mass order, i.e. $M \sim M_{Pl}$. In fact we can have the existence of extremal configurations with extremal mass $M_{\rm ext} \sim M_{Pl}$, along with a single degenerate horizon $r_{\rm ext}=r_-=r_+$.  By evaluating the solution numerically   it was found that $r_{\rm ext}=1.53231 [l_{\text{Pl}} ]$ and $ M_{\rm ext}=1.16537  [M_{\text{Pl}}]$. 
This case   plays an important role in the present paper.\\

(iii) Finally, there is the \textit{astrophysical black hole sector} in the region $M\gg M_{Pl}$, or the large mass limit. We will not focus on this case in the present work.

\section{Construction of the Entanglement-Induced Wormhole Geometry }
\label{SecIII}

In this section we construct the full spacetime geometry associated with an entangled two-particle system within the non-local gravitational framework introduced in the previous section. Starting from the regular metric supported by the gravitational self-energy, we show how an Einstein-Rosen–type bridge naturally emerges once the two entangled subsystems are embedded in this geometry. In particular, we will  analyze several distinct mass regimes, such as extremal particle–black-hole configurations, light two-particle states, and minimal-length geometries, and we will evaluate the resulting wormhole metrics, horizons, throat radii, embedding diagrams, and the appearance of exotic matter required at the minimal surface. Hence, this section  develops the complete geometric catalogue of wormholes produced by entanglement-induced non-local gravitational energy.

\subsection{Einstein-Rosen type wormhole}

Having established in the previous section the regular geometry and the non-local gravitational energy that governs the short-distance structure of spacetime, we now turn to the explicit construction of the wormhole sourced by two entangled subsystems. Inspired by the ${\rm ER}={\rm EPR}$ conjecture, let us construct an Einstein–Rosen wormhole geometry connecting two entangled particles.  As an example, we can consider a pair created in a maximally entangled state connected by an ER bridge with the spin state
\begin{equation}
\ket {\Psi} = \frac{1}{\sqrt{2}}\left(\ket{\uparrow_A \downarrow_B}+\ket{\downarrow_A \uparrow_B}\right).
\end{equation}
Therefore, it is natural to     propose a possible geometric interpretation in terms of a wormhole structure or the spacetime connecting entangled particles, analogous to the ${\rm ER}={\rm EPR}$ conjecture. At first, this seems to pose a problem, since one expects that it does not fit with entanglement, where you need to be able to go either way. However, it is known from the  \textit{no-communication theorem} that quantum entanglement does not allow sending information (or communication) but only in a sense of correlation between pairs. 



Returning to the ${\rm ER}={\rm EPR}$ conjecture, the key idea is that quantum correlations, and specifically entanglement, may have a 
geometric interpretation, suggesting a connection between entanglement and the structure of spacetime. To apply this idea, we first need to determine the spacetime geometry arising from our non-local theory of gravity, which was recently studied in  \cite{Jusufi:2025qgd}.
In particular, we   impose the spacetime metric \eqref{metric1}, introducing a new coordinate to facilitate this construction, namely
\begin{equation}
    u^2=r^2+l_0^2 \Longrightarrow u_{\pm}=\pm \sqrt{r^2+l_0^2},
\end{equation}
where  $u \in (-\infty, -u_{\rm min}] \cup [u_{\rm min}, \infty)$. 
The wormhole metric reads 
\begin{equation}\label{metric2}
    ds_{\pm}^2=-f(u_{\pm})dt^2+\frac{du_{\pm}^2}{\left(1-\frac{l_0^2}{u_{\pm}^2}\right) f(u_{\pm})}+(u^2_{\pm}-l_0^2)\,d\Omega^2,
\end{equation}
with
\begin{eqnarray}
    f(u_{\pm})&=&1-\frac{2M}{|u_{\pm}|}\left(1-\frac{l_0^2}{u^2_{\pm}}\right)+\frac{  M^2} {u_{\pm}^2} F(u_{\pm}),
\end{eqnarray}
where 
\begin{equation}
F(u_{\pm})=\frac{5}{8}-\frac{l_0^2}{4 u_{\pm}^2}-\frac{3 u_{\pm}^2}{8  l_0 \sqrt{u_{\pm}^2-l_0^2}}\arctan\left(\frac{\sqrt{u_{\pm}^2-l_0^2}}{l_0}\right).
\end{equation}

This metric is similar to the charge wormhole solution reported in \cite{Jusufi:2023dix}. In contrast, the spacetime \eqref{metric2} is purely neutral, and has not been reported before. We interpret $u_{\pm}$  mathematically as two congruent parts or ``sheets'', joined by a hyperplane at the wormhole throat. By construction, there should exist a smooth transition between the wormhole interior and the two external geometries. Later on we will point out an issue regarding the smoothness of the metric at the wormhole throat, which can be resolved by adding exotic matter at the throat.

To find the total mass, for large distances $|u_{\pm}| \gg l_0$, we can write 
\begin{equation}
    f(u_{\pm}) = 1-\frac{2  \left(M+\frac{3 \pi  M^2}{32 l_0  }\right)}{|u_{\pm}|}+\frac{ M^2}{u_{\pm}^2}\dots.\label{APPMETRIC}
\end{equation}
Therefore, for large distances $|u_{\pm}| \to \infty$, the Arnowitt–Deser–Misner (ADM) mass/energy reads
\begin{eqnarray}\label{ADM}
     \mathcal{M}= M+\frac{3 \pi M^2}{32l_0 },
\end{eqnarray}
which agrees with \cite{Jusufi:2025qgd}. Moreover, the surface area of the 2-sphere at fixed $u_{\pm}$ is:

\begin{equation}\label{surfa}
A = \int_0^{2\pi} \int_0^{\pi} \sqrt{g_{\theta\theta}g_{\phi\phi}} \, d\theta \, d\phi
  = 4\pi (u_{\pm}^2 - l_0^2).
\end{equation}
In constructing the wormhole geometry, we will consider two mass scales of specific wormhole configurations: the particle–black hole case with $r_{\min} = r_{\rm ext}$ having $M \sim M_{Pl}$, i.e. with black holes having Planck mass, and two examples of wormholes for the particle case with mass scale $M \ll M_{Pl}$ with $r_{\min} = \zeta l_0$ and $r_{\min} =0$, respectively. \\

\subsection{Entangled particle-black hole configuration} 

As we have seen, the scenario at hand
predicts particle-black hole objects with masses of the order of the Planck mass. 
Motivated by this, we aim to construct a corresponding wormhole spacetime. 
In particular, we consider another example of entangled particles that could give rise to wormhole formation: a pair of particle–black hole objects in an extremal configuration with maximal mass and radius. As we shall see, such a wormhole   has a horizon and it is one-way traversable.  
Such a spacetime possesses a horizon at $ r = r_{\text{ext}} $ and mass $ M = M_{\text{ext}}$. 

To proceed, we employ the extremal configuration, where for the minimal value we set
\begin{equation}
 r_{\rm min}=r_{\rm ext} \quad \text{and}\quad \,M=M_{\rm ext},
\end{equation}
and solving  the following conditions: $
f(r_{\rm ext},M_{\rm ext})=0 $ and $\left.  f'(r,M) \right|_{r = r_{\rm ext},M=M_{\rm ext}} = 0$, we get $r_{\rm ext}=1.53231 [l_0 ]$ and $M_{\rm ext}=1.16537 [M_{\text{Pl}}]$. For the minimal value of the new coordinate we obtain
\begin{equation}
 |u^{\rm min}_{\pm}|=\sqrt{r^2_{\rm ext}+l_0^2}= 1.8297 l_0,
\end{equation}
hence we get the interval  $u \in (-\infty, -1.8297 l_0] \cup [1.8297 l_0, \infty)$. 
This now implies that the determinant  of the metric is non-zero, i.e. $\det g_{\mu \nu}>0$, along with the wormhole throat which is now located at
\begin{eqnarray}
    r_{\rm throat}=\sqrt{ u_{\rm min}^2-l_0^2}=r_{\rm  ext}.
\end{eqnarray}
Using \eqref{surfa} we can compute the surface area of the 2-sphere at fixed $u_{\pm}$, which reads $A=4\pi r_{\rm  ext}^2$. 

At the throat of the wormhole, $ r_{\text{throat}} $, there exists a coordinate singularity (or horizon), 
which becomes evident when we set $ds^2 =0$ and fixed angles. From the metric we then obtain
\begin{equation}\label{dudt}
\left.\frac{du_{\pm}}{dt}= 
\pm f(u_{\pm}) \sqrt{1-\frac{l_0^2}{u_{\pm}^2}} \right|_{|u_{\pm}|=u_{\rm ext}}=0,
\end{equation}
since $f(u_{\pm})=0$ for the extremal case. This implies
that the wormhole has a horizon, and it is an one-way traversable wormhole with a vanishing Hawking temperature. In addition, since for this case the wormhole has a horizon, we can associate to it a  Bekenstein--Hawking entropy, which reads
\begin{eqnarray}
\label{Entropy_1}
    S_{BH}&= &\frac{k_B c^3 A }{ 4 G \hbar},
\end{eqnarray}
where $A=4 \pi r_{\rm throat}^2$ (restoring the international unit system). 
On the other hand, since $r_{\rm throat}=r_{ext} = 1.53231\, l_{0}$ 
(with $r_{ext}$  the cut through the ER bridge with minimal length)
and using $l_0 =\xi  l_{Pl}=\xi \sqrt{\hbar G/c^3}$,   in leading order terms and up to a factor of proportionality we obtain 
the entanglement entropy between a pair of particles, according to \cite{Dai:2020ffw} (but see also \cite{Jusufi:2023dix}) as
\begin{eqnarray}\label{Entropy_2b}
    S_{BH} = (1.53 \times \xi)^2 \frac{  \pi k_B c^3 l_{Pl}^2  }{G \hbar} \sim k_B \ln(2).
\end{eqnarray} 
Therefore, in order to have a complete correspondence we can set $\xi=\sqrt{\ln(2)/\pi}\times (1.53)^{-1}=0.3$.

\subsection{Entangled state of two particles}

The second case concerns an entangled state of two particles with mass scale $M \ll M_{Pl}$. Such a scenario is more plausible; for example, we can consider an $e^+ e^-$ pair created in a maximally entangled state, connected by an ER bridge. Since the particles can have charge such as the    
electron-positron case, one should add the effect of charge in the spacetime geometry. In fact, in \cite{Gaete:2022ukm}, in context of T-duality such a charged metric was derived and recently used in \cite{Jusufi:2023dix} in constructing a charged wormhole. We can add such an effect in our case and obtain the most general effect by combining the gravitational self energy and the electromagnetic field. 
In order to do this we can follow a similar approach as in the previous section, and
working with the radial coordinate we can calculate the effect of electromagnetic field as
\begin{eqnarray}
 E^{\rm EM}  &=& \frac{1}{8\pi}\int_0^r \textbf{E}^2(r') d^3\textbf{r}' \nonumber \\
&=&-\frac{5 Q^2 r^3}{16(r^2+l_0^2)^2}-\frac{3 Q^2 l_0^2 r}{16(r^2+l_0^2)^2}\nonumber \\
&&  +\frac{3  Q^2}{16 l_0} \arctan(\frac{r}{l_0}),
\end{eqnarray}
where $\textbf{E}$ is the electric field and in T-duality reads \cite{Gaete:2022une,Gaete:2022ukm}
\begin{eqnarray}
\textbf{E}=-\frac{Q r}{(r^2+l_0^2)^{3/2}}\hat{r}.
\end{eqnarray}
For the total energy density we have   
\begin{eqnarray}
    \rho=\rho^{\rm bare}+\rho^{\rm GSE}+\rho^{\rm EM},
\end{eqnarray}
and
in terms of the total mass
\begin{equation}
    m(r)=4 \pi \int_0^r \left[\rho^{\rm bare}(r') +\rho^{\rm GSE}(r')+\rho^{\rm EM}(r') \right]r'^2 dr'
\end{equation}
we extract the final solution
\begin{eqnarray}
    f(r)=1-\frac{2  M r^2}{ (r^2+l_0^2)^{3/2}}+\frac{  (M^2+Q^2) \,r^2  }{(r^2+l_0^2)^2} F(r)\label{metric1b},
\end{eqnarray}
where the function $F(r)$ is given in \eqref{F}.
This implies that the effect of charge is simply added by shifting the mass  
in the second term as $M^2\to M^2+Q^2$. Finally, we can construct the Einstein-Rosen type wormhole geometry  using $u_{\pm}=\pm \sqrt{r^2+l_0^2}$ with the wormhole metric (\ref{metric2}), which we rewrite here
\begin{equation}\label{metric2b}
    ds_{\pm}^2=-f(u_{\pm})dt^2+\frac{du_{\pm}^2}{\left(1-\frac{l_0^2}{u_{\pm}^2}\right) f(u_{\pm})}+(u^2_{\pm}-l_0^2)d\Omega^2,
\end{equation}
where
\begin{equation}
    f(u_{\pm})=1-\frac{2M}{|u_{\pm}|}\left(1-\frac{l_0^2}{u^2_{\pm}}\right)+\frac{  M^2+Q^2} {u_{\pm}^2} F(u_{\pm}).
\end{equation}
We normally expect the effect of mass to dominate the wormhole geometry, hence we will neglect the effect of charge in our elaboration by assuming $M \gg Q$.

\subsection{Specific wormhole subclasses}

We will elaborate two neutral wormhole examples: traversable/non-traversable wormholes
without a horizon and non-traversable wormholes with a horizon.

\subsubsection{Traversable/non-traversable wormholes without a horizon}
 
We construct an Einstein-Rosen type wormhole geometry between such particles using the choice 
\begin{equation}
r_{\rm min}= \zeta l_0,\quad \text{and} \quad M < M_{\rm ext},
\end{equation}
where $\zeta=1, 2, 3, ..$. In a sense, with the last equations we have imposed a quantization of the minimal area or, equivalently, the minimal throat radius of the wormhole geometry. In terms of the new coordinates, it yields 
\begin{equation}
|u^{\rm min}_{\pm}|=\sqrt{r^2_{\rm min}+l_0^2}=  l_0 \sqrt{\zeta^2+1}\, .
\end{equation}
Moreover, we have the following interval $u \in (-\infty, - l_0\sqrt{\zeta^2+1}] \cup [l_0 \sqrt{\zeta^2+1}, \infty)$. Similarly, the wormhole metric yields $\det g_{\mu \nu}>0$, along with the wormhole throat which is now located at
\begin{equation}
 r_{\rm throat}=\sqrt{ u_{\rm min}^2-l_0^2}= \zeta  l_0.
\end{equation}

In terms of  \eqref{surfa}, for each case for the surface area at the wormhole throat we obtain
\begin{equation}
 A=4 \pi  (u_{\rm min}^2-l_0^2)=4 \pi \zeta^2 l_0^2.
\end{equation}
 We see that at the throat of the wormhole $|u_{\pm}|=l_0 \sqrt{\zeta^2+1} $ there is no apparent/coordinate singularity. Furthermore, this choice leads to a regular and non-vanishing determinant of the metric, i.e. $\det(g_{\mu \nu}) = 0$. To see this, we can set $ds^2=0$, and from the metric we acquire 
 \begin{equation}
   \left.\frac{du_{\pm}}{dt} \right|_{|u_{\pm}|=l_0\sqrt{\zeta^2+1} }=\pm \frac{\zeta}{\sqrt{\zeta^2+1}}f(u_{\pm}),
\end{equation}
where, in all cases, $f(u_{\pm})\neq 0$ provided $M<M_{\rm ext}$. 
This wormhole has no horizon, and furthermore if we set $\zeta=1$
for the throat we obtain  $r_{\rm throat}=l_0$ while $|u^{\rm min}_{\pm}|=\sqrt{2}\, l_0$. 
Since $l_0$ denotes the minimal length, no real particle can pass through the wormhole throat.
Hence, for traversability we require at least $\zeta\ge 2$, which yields $r_{\rm throat} \geq 2 l_0$.

\subsubsection{Non-traversable wormholes with horizon}

The second example for an Einstein-Rosen type wormhole geometry for two entangled particles is a special case when $\zeta=0$, namely  
\begin{equation}
r_{\rm min}=0,\quad \text{and} \quad M < M_{\rm ext}.
\end{equation}
We can write this condition in terms of the new coordinates as $|u^{\rm min}_{\pm}|= l_0$, which
belongs to the interval $u \in (-\infty, - l_0] \cup [l_0,\infty)$. In this case the wormhole metric yields $\det g_{\mu \nu}=0$, along with zero wormhole throat $r_{\rm throat}=0$.
From  \eqref{surfa} we see that the surface area at the wormhole throat is zero, i.e.
$ A=4 \pi  (u_{\rm min}^2-l_0^2)=0.$ This implies
that this wormhole  geometry is not traversable. Furthermore, at the throat of the wormhole $|u_{\pm}|= l_0$ there is an apparent/coordinate singularity which can be seen by solving $ds^2=0$, namely
 \begin{equation}
 \left.\frac{du_{\pm}}{dt}= \pm f(u_{\pm}) \sqrt{1-\frac{l_0^2}{u_{\pm}^2}} \right|_{|u_{\pm}|=l_0}=0,
\end{equation}
where $f(u_{\pm})$ is regular for $|u_{\pm}|=l_0$. In summary, this 
wormhole represents a non-traversable wormhole with a horizon at the throat.

\subsection{Hawking temperature and radiation}

Let us now elaborate the Hawking temperature associated to the   horizons of the wormhole solutions of teh previous subsection.  To calculate the Hawking temperature we  
use a topological method based on the Gauss-Bonnet theorem  \cite{Robson:2018con,Robson:2019yzx}.
We rewrite the metric~\eqref{metric2} for the region $u=u_+$,
in a form of two-dimensional Euclidean spacetime having $\tau=i t$ and $d\Omega^2=0$, that is given by
\begin{equation}\label{2dE}
    ds^2=f(u)d\tau^2~+\frac{du^2}{f(u) h(u)},
\end{equation}
where we defined the function $h(u)=1-\frac{l_0^2}{u_{\pm}^2}$.
The Hawking temperature can be found from \cite{Robson:2018con} as
\be\label{Integral}
T_H=\frac{\hbar c}{4 \pi \chi k_B}\sum_{j \leq \chi}\int_{u_h} \sqrt{g}\, \mathcal{R}\, du~.
\ee

Applying this expression for the wormhole metric~\eqref{2dE}, we find the Ricci scalar 
\begin{equation}
\mathcal{R}=-h(u) f''(u)-\frac{h'(u)f'(u)}{2},
\end{equation}
and $\sqrt{g}=1/\sqrt{h(u)}$,  using the fact that the Euler characteristic of Euclidean geometry is $\chi=1$ at the wormhole horizon $u_h$.  We can apply this for two wormholes horizons (setting $\hbar=c=k_B=1$). 
The first one is for the extremal case $u_h=u_{ext}$ and the second one 
for zero-throat wormhole having $u_h=l_0$.  
Finally, we     solve the integral~\eqref{Integral}   obtaining
\begin{equation}
T_H=\frac{1}{4 \pi}\int_{u_h}^{\infty} \mathcal{R} \sqrt{g}\, du=-\left.\frac{\sqrt{u^2-l_0^2}}{u} f'(u) \right|_{u_h}^{\infty}~.
\end{equation}

For the extreme wormhole case we know that $f(u_{ext})=f'(u_{ext})=0$ vanishes, while   asymptotically
the solution has the form of   \eqref{APPMETRIC}, and thus  we get $\lim_{u \to \infty}f(u)=1$ yielding $f'(u)=0$ as $u \to \infty$. This results to $T_H=0$, as expected. Lastly, from the thermodynamical point of view this exteremal wormhole is stable. 

For the second case having $u_h=l_0$, it can  be easily seen that the first term   vanishes, i.e.
$\lim_{u \to l_0} \sqrt{u^2-l_0^2}=0$,  and moreover $f'(u)=0$ as $u \to \infty$ at asymptotic infinity, again giving
a vanishing Hawking temperature. We conclude that this wormhole is also thermodynamicaly stable. Note that the same conclusions can be obtained using a standard approach in calculating Hawking temperature.

\subsection{Embedding diagrams}

Let us now analyze in detail the embedding diagrams, since these are useful to impose the requirement that 
the spacetime metric (\ref{metric2}) describes a wormhole geometry. Of particular interest is the 
equatorial slice $\theta = \pi/2$ at a fixed moment in time, $t = \mathrm{const}$. 
Under these constraints, the metric~(\ref{metric2}) reduces to
\begin{equation}\label{u-metric_emb1}
    ds^2 = \frac{du_{\pm}^2}{\left(1 - \frac{l_0^2}{u_{\pm}^2}\right) f(u_{\pm})} 
    + \left(u_{\pm}^2 - l_0^2\right)d\phi^2.
\end{equation}
We can further compute the proper length or distance using
\begin{eqnarray}
    l=\pm \int_{u_{min}}^{\pm \infty} \frac{du_{\pm}}{\sqrt{\left(1-\frac{l_0^2}{u_{\pm}^2}\right) f(u_{\pm})}}.
\end{eqnarray}

This reduced metric can be embedded into a three-dimensional Euclidean space, which in cylindrical 
coordinates $(r, \phi, z)$ takes the form
\begin{equation}\label{u-metric_emb2}
    ds^2 = dz^2 + du_{\pm}^2 + \left(u_{\pm}^2 - l_0^2\right)d\phi^2.
\end{equation}
The embedded surface $z(u_{\pm})$ can then be obtained by comparing the last two line elements and 
solving for the slope:
\begin{equation}
    \frac{dz}{du_{\pm}} = \pm 
    \sqrt{\frac{1}{\left(1 - \frac{l_0^2}{u_{\pm}^2}\right) f(u_{\pm})} - 1}.
\end{equation}

\begin{figure}[ht!]
\centering
\includegraphics[scale=0.21]{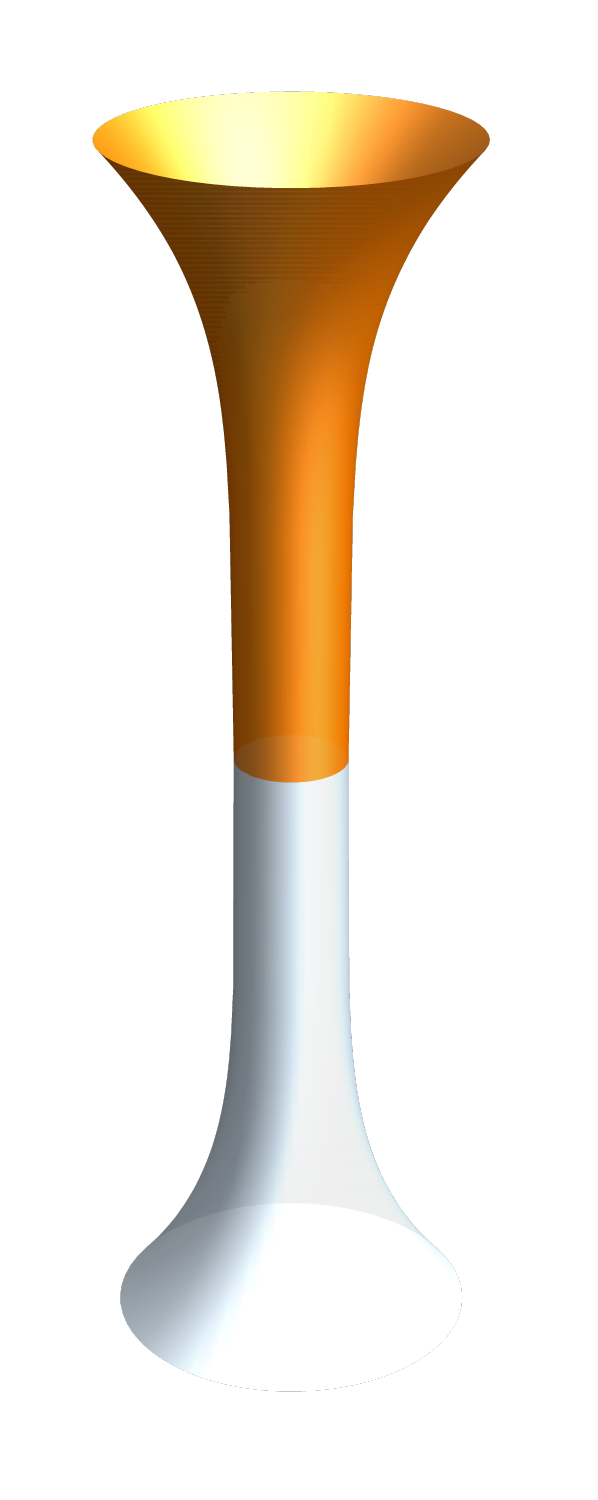}
\caption{{\it{Embedding diagram of the one-way traversable wormhole with horizon (extremal configuration). This represents an entanglement-induced wormhole geometry between two entangled particle–black holes of Planck-mass order. We have used $u_{\min}=1.8297$ and $M_{\text{ext}}=1.16537$, and we have set the Planck quantities to unity, namely $M_{\text{Pl}}=l_{0}=1$.}}}
\label{fig:wormhole1}
\end{figure}

\begin{figure}[ht!]
\centering
\includegraphics[scale=0.7]{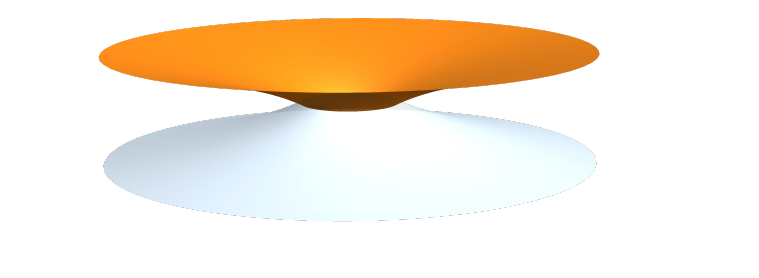}
\caption{{\it{Embedding diagram of the two-way traversable wormhole without a horizon for the case $\zeta=2$ which implies that $u_{\min} = \sqrt{5}$. This represents an entanglement-induced wormhole geometry between two entangled particles of particle mass sector.  We have set $M = 0.1$ along with $M_{\text{Pl}}=l_{0}=1$.}}}
\label{fig:wormhole2}
\end{figure}

\begin{figure}[ht!]
\centering
\includegraphics[scale=0.21]{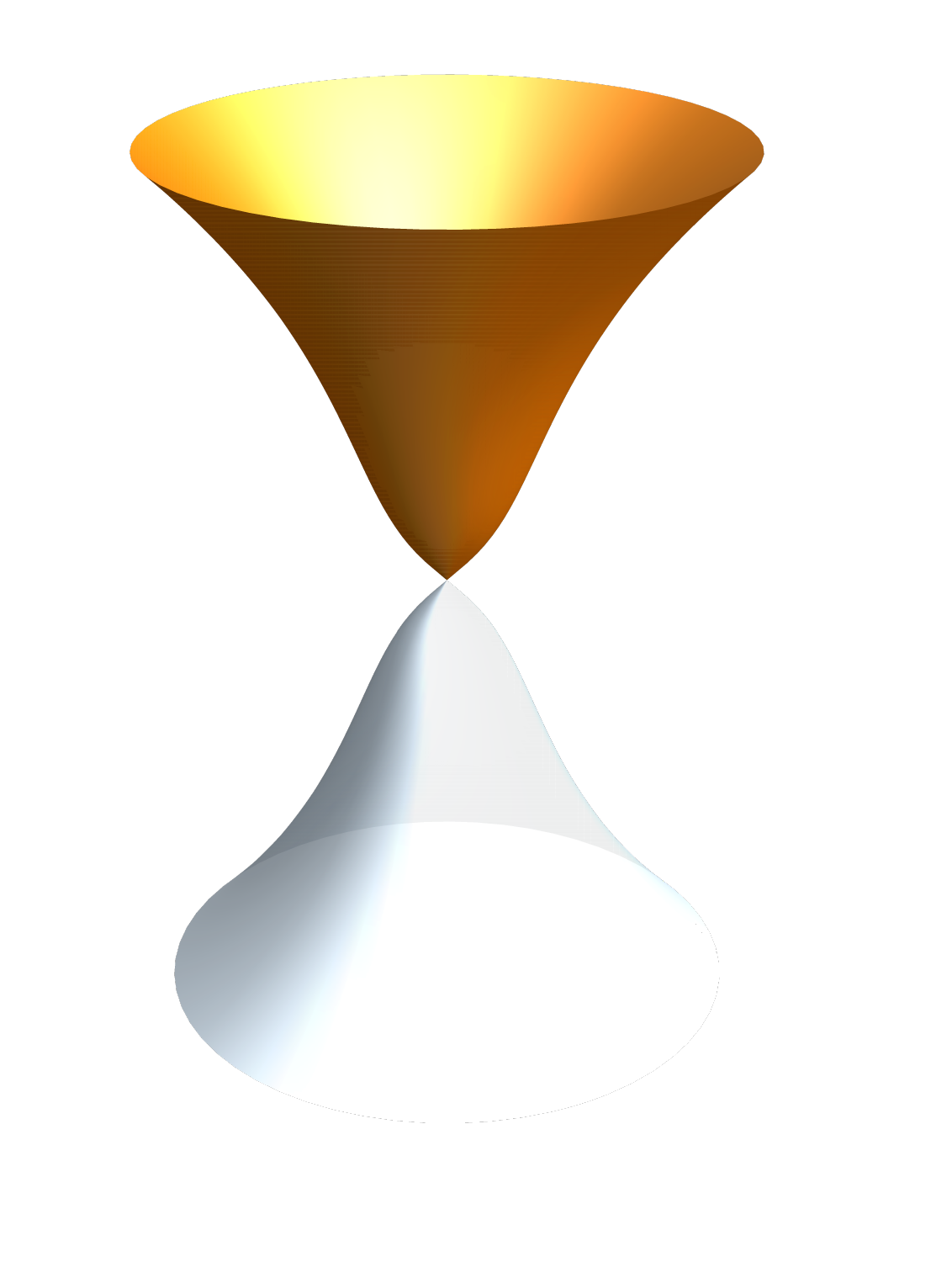}
\caption{{\it{Embedding diagram of the non-traversable wormhole with horizon having zero throat radius $r_{\rm throat}=0$ and $u_{\min} =1$, respectively. This also represents an entanglement-induced wormhole geometry between two entangled particles of particle mass sector. We have further set the mass parameter to $M = 0.1$, along with  $M_{\text{Pl}}=l_{0}=1$.}}}
\label{fig:wormhole3}
\end{figure}

In Figs.~\ref{fig:wormhole1}-\ref{fig:wormhole3} we present the embedding diagrams for three types of wormhole geometries. The first corresponds to a spacetime generated by two entangled particle-black holes, 
and the second and   third correspond  to a spacetime generated by two entangled particles. 
In the first case the wormhole has a horizon, and we can further say that this spacetime is an one-way traversable wormhole. On the other hand, 
in the second case the wormhole has no horizon, and   in principle this spacetime 
is a two-way traversable wormhole, while in the third case 
the wormhole  again has a horizon and it is non-traversable. Note that a significant difference in their geometrical structure can be observed by changing the mass scale.\\

\subsection{Exotic matter}

We proceed by  elaborating further  the wormhole geometry and examine
the presence of exotic matter at the throat. Recall that exotic matter is defined as matter that violates the null energy condition \cite{Morris:1988cz,Visser:1995cc,Lobo:2017cay}.
The metric \eqref{metric2} is smooth everywhere apart from 
the wormhole throat located at $u_{\min}$. In fact, it does not satisfy the Einstein field equations there, requiring a source of exotic matter. The issue of smoothness for the original Einstein–Rosen metric was discussed in \cite{Guendelman:2009er,Guendelman:2009zz}, and more recently in the context of the BTZ wormhole in \cite{Anand:2024vmy}. 

To support our argument, we use the Levi-Civita identity \cite{Guendelman:2009er,Guendelman:2009zz}
\begin{eqnarray}
    R^t_t=-\frac{1}{\sqrt{-g_{tt}}}\nabla^2_{(3)}\left(  \sqrt{-g_{tt}} \right) \ ,
\end{eqnarray}
where $\nabla^2_{(3)}$ is the three-dimensional spatial Laplacian. The metric \eqref{metric2} solves the Einstein field equations for all $|u_{\pm}| \neq u_{min}$, however, at $|u_{\pm}|=u_{\rm min}$ the metric is not smooth, that is, there is a jump and a matter source is needed. 
This can be seen from the fact that the  involved  equation contains second-order derivative terms like
\begin{equation}
    \frac{\partial^2 }{\partial u_{\pm}^2} \sqrt{-g_{tt}} \sim \frac{\partial^2 }{\partial u_{\pm}^2} |u_{\pm}| \sim  2\delta (u_{\pm}) \ .
    \label{smooth}
\end{equation} 
As shown above, the metric is smooth everywhere apart from  the minimal surface 
$u = u_{\min}$, where the second derivative of $\sqrt{-g_{tt}}$ is discontinuous, namely  $\frac{\partial^2}{\partial u^2}\sqrt{-g_{tt}}
	\;\sim\; 2\,\delta(u-u_{\min}) \, .$
This discontinuity implies the presence of a thin layer of matter localized at the throat, described by a surface stress--energy tensor
\begin{equation}
	S^{a}{}_{b} = -\frac{1}{8\pi}\left([K^{a}{}_{b}] - \delta^{a}{}_{b}[K]\right),
\end{equation}
where $K_{ab}$ is the extrinsic curvature of the constant-$u$ hypersurface and $[X]$ denotes the jump across $u = u_{\min}$.

For the metric~(\ref{metric2}), the induced metric on the throat is 
\begin{equation}
	h_{ab} = \mathrm{diag}\left(-f(u_{\min}),\, u_{\min}^2 - l_0^2,\, u_{\min}^2 - l_0^2 \right).
\end{equation}
A short calculation yields the surface energy density
\begin{equation}
	\sigma = -\frac{1}{4\pi}\frac{\sqrt{f(u_{\min})}}{u_{\min}} \,,
\end{equation}
and the transverse pressures
\begin{equation}
	p_T = \frac{1}{8\pi}\left[ \frac{\partial_u f(u)}{\sqrt{f(u)}} \right]_{u=u_{\min}} .
\end{equation}
Hence, the total amount of exotic matter required at the throat is  
\begin{eqnarray}
	\Omega_{\rm exotic}
	&=& \int \sigma \, dA
	= 4\pi (u_{\min}^2 - l_0^2)\, \sigma
	\nonumber\\
	&=& - (u_{\min}^2 - l_0^2)\frac{\sqrt{f(u_{\min})}}{u_{\min}} .
\end{eqnarray}

For microscopic throats, $u_{\min} \sim l_0$, the exotic matter content scales as
\begin{equation}
	\Omega_{\rm exotic} \sim - \mathcal{O}(l_0),
\end{equation}
which is of Planck-scale magnitude and therefore finite and extremely small. This is consistent with the fact that the spacetime is regular, and exotic matter appears only as a localized thin layer induced by quantum gravitational effects.
In what follows, we shall elaborate in more detail on some of the implications of our results in the context of the ER = EPR conjecture.

\section{Physical Interpretation and ER=EPR Constraints}\label{SecIV}

Having obtained the full set of wormhole geometries supported by the non-local gravitational energy, we now examine which of these configurations can be consistently interpreted as entanglement-induced Einstein–Rosen bridges.
In this section we examine the quantum-gravity origin of exotic matter, evaluate the causal and non-traversability conditions required for a wormhole to represent an entangled EPR pair, and determine which of the geometries obtained in Sec. \ref{SecIII} satisfy these constraints. We further explore broader implications, such as microscopic ER networks in the quantum vacuum and wormholes associated with Hawking-pair entanglement, to show
how the emergent wormhole structures may reflect the connection between entanglement and spacetime.

\subsection{Quantum-gravity-induced exotic matter and black-hole--to--wormhole transitions}

Both the bare energy density $\rho^{\rm bare}$ and the non-local gravitational self-energy density $\rho^{\rm GSE}$ violate the SEC in a compact region 
\begin{equation}
	|u| \lesssim \sqrt{\frac{5}{3}}\,l_0,
\end{equation}
due to the regularizing zero-point length. This violation arises naturally from the quantum structure of spacetime and does not require an ad hoc choice of stress--energy sources.

Regular black holes constructed in this framework   contain a built-in region where the SEC fails. Such a region provides precisely the exotic stress--energy required to support a wormhole throat. Consequently, when two regular black holes are quantum-mechanically entangled, it becomes plausible for the correlated cores to be connected by a non-traversable wormhole, in accordance with the ${\rm ER}={\rm EPR}$ correspondence.
In contrast, classical singular black holes do not possess an interior region capable of sourcing exotic matter. Hence, the ${\rm ER}={\rm EPR}$ conjecture is realized more naturally  for regular black holes than for their singular counterparts.

\subsection{ER=EPR constraints}

The ${\rm ER}={\rm EPR}$ correspondence requires that any geometric realization of quantum entanglement must satisfy several core physical constraints. These arise from fundamental features of quantum theory and general relativity, in particular the no-communication theorem, causal structure, and the properties of non-traversable wormholes. The purpose of this subsection is to analyze these requirements in detail and to determine which wormhole geometries obtained in our construction are compatible with the ${\rm ER}={\rm EPR}$ framework.

Quantum entanglement correlates spatially separated subsystems, however it does not allow communication between them. The no-communication theorem states that no choice of local measurement performed by one observer can affect the statistics observed by another observer. Any spacetime interpretation of entanglement must therefore prohibit the transfer of signals or information across the corresponding Einstein-Rosen bridge. A traversable wormhole would violate this principle, since even a very small region in which null geodesics could propagate from one exterior region to the other, would enable superluminal or acausal communication. Consequently, a physically acceptable ER bridge must be non-traversable for both lightlike and timelike trajectories.

 On the other hand, non-traversability can arise for two reasons. Either the wormhole geometry contains an event horizon that causally disconnects the two exterior regions, or the throat radius is too small for classical or quantum fields to penetrate. The first mechanism is the conventional method by which non-traversability is ensured in classical GR. A horizon prevents any signal from crossing the bridge, and the induced causal structure agrees with the standard interpretation of the EPR correlations. The presence of a horizon is furthermore consistent with the original Einstein–Rosen construction, where the bridge appears as a non-traversable wormhole created from the maximal extension of the Schwarzschild spacetime. Any candidate geometry for ${\rm ER}={\rm EPR}$ must reproduce this causal feature.
 
In classical GR, black holes contain curvature singularities which obstruct any smooth continuation into an interior region that could support a wormhole throat. The regular geometries obtained in non-local gravity resolve these singularities and introduce a minimal length scale $l_0$. This length scale plays a dual role. Firstly, it ensures that the geometry remains finite at the core, preventing divergences in curvature invariants. Secondly, 
it produces quantum-gravity-induced violations of the SEC, which supply the exotic matter required to sustain a wormhole throat. This suggests that regular black holes are more natural candidates for ER bridges than their singular counterparts.
 
The solutions presented in Sec. \ref{SecIII}  produce three families of wormhole geometries. Each one must be assessed with respect to the two ${\rm ER}={\rm EPR}$ requirements: (i) non-traversability and (ii) causal horizons. 
 
\begin{enumerate}
	\item \textit{Particle–black-hole wormhole (extremal case).}  
	This geometry contains an event horizon located at the minimal radius $r_{\rm ext}$, thus satisfying the causal requirement. However, the wormhole is one-way traversable in the sense that infalling observers can cross the horizon and reach the interior, although no signals can return. This partial traversability is not entirely compatible with ${\rm ER}={\rm EPR}$ since the presence of a one-way path is in tension with strict non-communication. Therefore, this geometry is only marginally acceptable.
	\item \textit{Two-particle wormholes with $\zeta \ge 2$.}  
	These spacetimes exhibit a smooth throat of finite non-zero radius and have no horizon. Lightlike and timelike geodesics can propagate across the throat in both directions. This creates a two-way traversable wormhole. Such geometries directly conflict with the no-communication theorem. As a result, they cannot represent ER bridges.
	\item \textit{Minimal-throat wormhole ($r_{\rm throat}=l_0$).}  
	The throat radius equals the fundamental minimal length scale $l_0$. No classical particle with finite rest mass can traverse such a throat, since any localized wavepacket cannot be compressed below this scale. The spacetime is horizonless, yet it remains physically non-traversable for realistic matter. This geometry satisfies the non-traversability requirement but does not satisfy the horizon requirement. It is therefore partially compatible with ${\rm ER}={\rm EPR}$.
	\item \textit{Zero-throat wormhole ($r_{\rm throat}=0$, $M \ll M_{\rm Pl}$).}  
	In this configuration the throat area vanishes and an apparent horizon forms at $u=l_0$. No null or timelike geodesic can cross from one exterior region to the other. The spacetime is fully regular and the causal structure precisely matches the requirements of the ${\rm ER}={\rm EPR}$ framework. It possesses a horizon, satisfies strict non-traversability for both light and massive particles, and is supported by quantum-gravity-induced exotic matter in a finite region. Hence, only the zero-throat configuration simultaneously satisfies all causal and geometric requirements of ${\rm ER}={\rm EPR}$. This solution is therefore the most physically consistent candidate for a geometric realization of entanglement. 
     
\end{enumerate}
 
 The wormhole configurations considered in our construction are, to some extent, quantum–mechanical objects. As such, systems with higher mass are expected to be unstable, while systems with very small mass can be stable for longer periods of time. This is a result of instabilities induced by self–gravitational effects.  
The characteristic timescale of an instability can be estimated using the uncertainty relation
$
\tau \sim \frac{\hbar}{\Delta E},
$
where $\Delta E$ denotes the energy uncertainty. 
One possible estimate for $\Delta E$ follows from (\ref{APPMETRIC}) which can be written as
$
f(u_{\pm})=1- \frac{2 E(u_{\pm})}{u_{\pm}}+\dots,
$
where the energy is defined as 
$
E(u_{\pm})=\mathcal{M}-\frac{M^2}{2u_{\pm}},
$
where total mass $\mathcal{M}$ is given by Eq. \eqref{ADM}.  From this last equation, we can get the uncertainty in energy as
$
\Delta E \sim \frac{M^2}{2u_{\pm}^2} {\Delta u_{\pm}}.
$
If we assume that the uncertainty in distance is of the order of the distance, i.e., $\Delta u_{\pm} \sim u_{\pm}$ and by restoring the physical constants $\hbar$, $c$ and $G$, for the instability timescale of the wormhole we obtain
\begin{equation}
\tau \sim \frac{\hbar\, M_{\rm Pl} \, u_{\pm}}{M^{2} l_{\rm Pl} c^{2}}.
\end{equation}

Thus, we found that the instability time of the wormhole is strongly dependent on the mass of the system. This time also depends on the distance $u_{\pm}$ and, in general, for the uncertainty distance  we can have $u_{\pm} \gg l_{\rm Pl}$. As a special case, we can have two entangled particles located at some minimal uncertainty distance $u_{\pm} \sim l_{\rm Pl}$, and further if we take Planck mass wormhole with $M \sim M_{\rm Pl}$, the wormhole collapses extremely rapidly. However, for smaller masses $M \ll M_{\rm Pl}$, the system can exist for much longer times. Therefore, two entangled objects with large masses would correspond to a wormhole connection that decays almost instantaneously, while for small masses and imposing $ u_{\pm} \gg l_{\rm Pl}$, the connection can exist for a long period of time. For example, if we take the distance $ u_{\pm} \sim 10^{-7}$ m and $M \sim 10^{-30}$ kg we get $\tau \sim 10^{30}$ s. Although such a small mass wormhole can exist for a long period of time, in reality it will eventually collapse within a short timescale due to environmental effects.

Combining the physical and geometric criteria, the zero-throat wormhole with mass $M \ll M_{\rm Pl}$ emerges as the preferred ER bridge geometry in the present model. Moreover, it satisfies all constraints required by the ${\rm ER}={\rm EPR}$ correspondence: it is non-traversable and regular, it possesses a horizon, it avoids causal paradoxes, and it naturally incorporates exotic matter through quantum gravity effects. Other wormhole geometries produced by the construction
satisfy only a subset of these requirements. Therefore, the zero-throat configuration  provides the most coherent and physically admissible realization of an Einstein–Rosen bridge associated with entangled quantum states.

\subsection{Implications}

The above analysis  identifies the zero-throat wormhole geometry as the most physically admissible realization of an ER bridge, within our construction.
While this result is primarily theoretical, it naturally leads to two speculative but conceptually appealing implications regarding the microscopic structure of spacetime, and possible connections to dark-energy phenomenology and Hawking radiation.

\paragraph{Quantum-vacuum fluctuations and microscopic ER networks.}
The quantum vacuum is not an empty background but a highly dynamical state characterized by fluctuations of quantum fields. Virtual particle pairs are continuously created and annihilated, and within the ${\rm ER}={\rm EPR}$ framework \cite{maldacena2013cool} any entangled pair can, in principle, be associated with a non-traversable Einstein--Rosen bridge. In the context of our model, the zero-throat wormhole configuration, with $M \ll M_{\rm Pl}$ and a horizon located at the minimal length scale $l_0$, provides a concrete geometric realization of such microscopic bridges.

If vacuum fluctuations continually generate entangled virtual particles, the resulting spacetime foam may contain an immense network of Planck-scale, horizon-supported, non-traversable wormholes. Since the entanglement entropy associated with these configurations scales holographically,
\begin{equation}
	S_{\rm ent} \propto \frac{R_H^2}{\ell_{\rm Pl}^2},
\end{equation}
a sufficiently large ensemble of microscopic ER bridges could, in principle, reproduce an entropy of order
\begin{equation}
	S_{\rm DE} \sim 10^{122} k_B,
\end{equation}
consistent with the holographic entropy associated with dark energy~\cite{Lee:2018lgs}. While this idea is speculative, it offers a natural mechanism within our framework for generating a macroscopic entanglement entropy from microscopic wormhole structures, in line with previous suggestions in the literature~\cite{Licata:2025rik,Jusufi:2023dix,Tsilioukas:2023tdw,Tsilioukas:2024seh,Anagnostopoulos:2025tax,Petronikolou:2025mlm}.

\paragraph{Hawking radiation and entangled wormhole pairs.}
A second speculative but physically motivated realization of the ${\rm ER}={\rm EPR}$ idea arises in the context of Hawking radiation~\cite{Hawking:1975vcx}. When a black hole emits a Hawking quantum, the outgoing particle remains 
quantum-mechanically entangled with its interior partner. Geometrically, this entanglement can be described as a tiny, non-traversable wormhole connecting the two quanta, consistent with the ${\rm ER}={\rm EPR}$ correspondence~\cite{maldacena2013cool}.

This interpretation is further supported by modern developments in the gravitational path-integral approach. In particular, replica wormholes~\cite{Penington:2019kki,Almheiri:2019qdq} have been shown to be essential for reproducing the Page curve~\cite{Page:1993wv,Page:1993df}, thereby demonstrating how information becomes encoded in the Hawking radiation. These results suggest that wormhole-like structures emerge in the semiclassical expansion of the gravitational path integral.

Within our construction, the zero-throat wormhole again provides a natural geometric candidate for such microscopic connections. Each pair of entangled Hawking quanta could correspond to a minimal-length non-traversable ER bridge, providing a simple and concrete realization of the ${\rm ER}={\rm EPR}$ correspondence during the evaporation process. Although idealized, this viewpoint aligns with contemporary approaches to the black-hole information problem and offers a possible microscopic origin for the wormhole structures appearing in replica-based entropy calculations. 

The aforementioned two speculative scenarios, namely the 
vacuum network of Planck-scale ER bridges and the Hawking-pair wormholes, illustrate how the geometries derived in our construction may have broader implications for understanding spacetime microstructure, entanglement entropy, and the quantum-gravitational nature of correlations. While preliminary, both ideas represent natural extensions of the ${\rm ER}={\rm EPR}$ correspondence once the zero-throat wormhole is adopted as the fundamental geometric element associated with quantum entanglement.

\section{Conclusions}\label{SecV}

In this work, we applied concepts from T-duality and incorporated the effects of non-local gravitational energy stored in the field into the wormhole spacetime geometry. 
We constructed an Einstein-Rosen wormhole geometry for  entangled states, involving two particles or two particle–black holes. We argued that at the throat of the wormhole, matter with negative energy is required to keep the wormhole open. In fact, even for regular spacetimes the Strong Energy Condition is violated which impies that when two entangled particles form a wormhole, such exotic energy at the wormhole throat is required to stabilize the
geometry, i.e. to keep the wormhole throat open. This result follows from the fact that ${\rm ER}={\rm EPR}$ strictly requires a non-traversable bridge with no macroscopic throat, so that no causal signal can pass between the entangled subsystems.

Compared to the well-known Einstein–Rosen metric, which is derived from the singular Schwarzschild solution and thus fails to describe the particle sector and the need for exotic matter, our wormhole construction is based on a regular geometry that consistently describes both the particle and black hole sectors. In this case,  the presence of exotic matter, linked to the violation of energy conditions, emerges naturally from quantum gravity effects in regular spacetimes, allowing for a smooth black hole-wormhole transition between two entangled states. 
The full dynamics of this transition is beyond the scope of this work. 
On one hand, this perspective provides a potential argument in support of the ${\rm ER}={\rm EPR}$ conjecture \cite{maldacena2013cool}, while on the other hand it offers an argument that quantum entanglement and spacetime are intimately related~\cite{VanRaamsdonk:2010pw}. 

Such wormholes should be, to a large extent, quantum-mechanical objects, and by using the time-energy uncertainty relation we estimated the possible timescale of these wormholes, which depends on the mass of the system. Specifically, Planck-mass wormholes would collapse rapidly due to self-gravitational effects, while low-mass wormholes (such as those associated with elementary particles) could exist for a long period of time. However, even these wormholes are expected to collapse within a short timescale due to environmental effects.

Finally, we speculated that our wormhole models can contribute in the form of entanglement entropy in the universe, and thus they can be applied to describe the dark energy from the vacuum fluctuations. The quantum fluctuations of particles can lead to a network of non-traversable Einstein–Rosen bridges. In addition, such wormholes can play important role in the context of Hawking radiation.  Namely, Hawking quanta can be quantum-entangled and under the ER = EPR conjecture, this entanglement can be reinterpreted geometrically as a microscopic non-traversable wormhole connecting the two particles via the replica wormholes connecting in this way the interior of the black hole and the Hawking radiation in the exterior of the black hole.


The framework developed in the present manuscript opens the possibility of studying fully dynamical and backreacting wormhole formation in evolving entangled systems, including time-dependent entanglement entropy, decoherence, or particle creation in curved spacetime. Extending our analysis to multi-particle or network-like entanglement patterns, where non-local gravitational energy could induce a web of microscopic ER bridges, may illuminate their role in spacetime foam or holographic tensor-network models, as well as in higher-spin, charged, or supersymmetric sectors. Embedding the construction in a semiclassical or effective-field-theory framework, combined with numerical simulations of the wormhole--particle--black-hole dynamics, could clarify the interplay between non-local gravitational energy, quantum information flow, and black-hole evaporation, enabling a quantitative characterization of the wormhole stability under realistic perturbations. Confronting these developments with holographic and quantum-information–inspired approaches would help elucidate how quantum correlations affect spacetime geometry and how ${\rm ER}={\rm EPR}$ manifests beyond the static two-particle sector, advancing a deeper understanding of the geometric nature of entanglement and its role in shaping the microscopic structure of spacetime.

\begin{acknowledgments}
FSNL acknowledges funding from the Fundacão para a Ciência e a Tecnologia (FCT) through the research grants UIDB/04434/2020, UIDP/04434/2020 and PTDC/FIS-AST/0054/2021. FSNL also acknowledges support from the FCT Scientific Employment Stimulus contract with reference CEECINST/00032/2018.  ENS   acknowledges  the 
contribution of the LISA Cosmology Working Group (CosWG).
KJ, FSNL and  ENS gratefully acknowledge  the   support from the COST 
Actions CA21136 -  Addressing observational tensions in cosmology with systematics and fundamental physics (CosmoVerse)  - CA23130, Bridging high and low energies in search of quantum gravity (BridgeQG) and CA21106 -- COSMIC WISPers in the Dark Universe: Theory, astrophysics and experiments (CosmicWISPers). DS acknowledges a CSU Fresno RSCA award and the Frank Sutton Research Fund for support.
\end{acknowledgments}

\bibliography{ref}

\end{document}